\documentclass[showpacs,twocolumn,prb]{revtex4-1}
\usepackage{amsfonts}
\usepackage{graphicx}
\usepackage{amsmath}
\usepackage{amssymb}

\begin{document}

\title{Electron-phonon correlations on spin texture of gapped helical Dirac
Fermions}

\author{Zhou Li$^{1,3,4}$}

\email{lizhou@mcmaster.ca}

\author{J. P. Carbotte$^{1,2}$}

\email{carbotte@mcmaster.ca}

\affiliation{$^{1}$ Department of Physics, McMaster University,
Hamilton, Ontario Ontario, Canada,L8S 4M1 \\
 $^{2}$ Canadian Institute for Advanced Research, Toronto, Ontario,
Canada M5G 1Z8 \\
 $^{3}$ Department of Physics and Astronomy, Louisiana State University,
Baton Rouge, LA, 70803 USA \\
 $^{4}$ Center for Computation and Technology, Louisiana State University,
Baton Rouge, LA, 70803 USA} 
\begin{abstract}
The metallic surface states of a topological insulator support helical
Dirac fermions protected by topology with their spin locked perpendicular
to their momentum. They can acquire mass through magnetic doping or
through hybridization of states on opposite faces of a thin sample.
In this case there can be a component of electron spin oriented perpendicular
to the surface plane. The electron-phonon interaction renormalizes
the dynamics of the charge carriers through their spectral density.
It also modifies the gap channel and a second spectral function enters
which, not only determines the out of plane spin component, but also
comes into in-plane properties. While the out of plane spin component is decreased below the Fermi momentum ($k_F$), 
the in plane component increases. There are also correlation tails extending well beyond $k_F$. The angular resolved 
photo-emission line shapes aquire Holstein side bands. The effective gap in the density of states is reduced and the optical
conductivity aquires distinct measurable phonon structure even for modest value of the electron-phonon coupling.

\end{abstract}

\date{\today }

\maketitle
\section{Introduction}

Helical Dirac fermions protected by topology and found at the surface
of topological insulators (TI)\cite{Hasan,Qi1,Moore,Hsieh1} exhibit
spin locked perpendicular to their momentum.\cite{Chen1,Hsieh1,Hsieh2,Jozwiak,Xu}
Doping with magnetic impurities can break time reversal symmetry and
create massive Dirac fermions as has been experimentally demonstrated\cite{Chen2}
in $Bi_{2}Se_{3}$. A gap can also be introduced in thin films when
the distance between top and bottom surfaces is of the order of the
extend in space of the surface states. \cite{Lu,Linden} Gapped Dirac fermions
arise in many other systems, for example in two dimensional membranes
such as the dichalcoginide $MoS_{2}$ \cite{Mak,Splend,Lee,Feng,Li1}and
silicene\cite{Aufray,Padova,Stille,Ezawa1,Ezawa2} with buckled honeycomb
lattice. In both these cases pseudospin plays the analogous role to
the real spin of topological insulators.

The electron-phonon interaction renormalizes quasiparticle dynamics
and leads to important observable changes in electronic properties
\cite{Carbotte,Li2,Li3,Li4,Stauber,Pound2,Ion,Leblanc1,Leblanc2,Pound3}
which illustrate the effects of many body renormalizations not captured
in single particle theories. For the Dirac electrons in graphene as
an example, features observed in the density of states \cite{Pound3,Miller,GLi,Nicol}
and in the dispersion curves measured in angular resolved photo emission
spectroscopy \cite{Zhou} have been interpreted as phonon structure.
\cite{Pound}The optical properties of graphene are also renormalized
in a non trivial way.\cite{Carbotte,Li4,Stauber,zli,Carbotte1} In
a simple bare band picture there is no optical absorption in the photon
energy region between the Drude intraband contribution centered about
$\omega=0$ and the interband onset at twice the value of the chemical
potential $\omega=2\mu$. In reality the real part of the dynamic
longitudinal conductivity is observed to be finite and almost one
third its universal background\cite{Gusynin1,Gusynin2} value in this
photon region.\cite{zli} This absorption is due to many body renormalizations
and is at least partially assigned \cite{Carbotte,Stauber} to the
electron-phonon interaction which provides boson assisted processes
referred to as Holstein processes. A phonon is created along with
an electron-hole pair.

Understanding the transport properties of the Dirac electrons on the surface of 
a topological insulator is important for possible device applications. 
At finite temperature the electron-phonon interaction is expected to be an 
important scattering chanel \cite{Hatch}, X. Zhu et al. \cite{Zhu1} have studied the surface phonons
on the (001) surface of $Bi_{2}Se_3$ and in particular have found a giant Kohn anomaly
associated with a branch having a maximum of 18 THz. From measurements of the phonon self energy, 
the same group \cite{Zhu2} determined the size of 
the electron-phonon interaction report a coupling constant of 0.43 for a particular branch which is 
much larger than reported in angular resolved photo-emission spectroscopy. One such study 
by Z. H. Pan et al. \cite{Pan} gives a mass enhancement $\lambda$ of 0.08 
while another by R. L. Hatch et al. \cite{Hatch} found $\lambda\sim$ 0.25. 
A more recent report \cite{Chen} gives $\sim$ 0.17 with characteristic phonon energy $\sim$18 meV.

The results of X. Zhu et al. are further supported by an infrared study by 
A. D. LaForge et al. \cite{LaForge} which found a strong electron-phonon 
coupling to a 7.6 meV optical phonon while S. Giraud et al. \cite{Giraud1}
provide arguments for coupling to acoustic phonon with $\lambda\sim$ 0.42 in their films and even 
larger in other geometries. \cite{Giraud2} 
Very recently J. A. Sobota et al. \cite{Sobota}, using time-resolved photoemission 
spectroscopy, find evidence for coupling to a 2.05 ThZ surface phonon mode with the Dirac 
electrons in $Bi_{2}Se_3$. Possible complications in the interpretation of optical 
pumping on time resolved data were discussed by S. Ulstrup \cite{Ulstrup} 
and need to be kept in mind. Finally, the recent transport measurements 
of M.V. Costache et al. \cite{Costache} were interpreted 
with strong coupling to a single optical phonon mode of energy $\sim$ 8 meV.

In this paper we will emphasize the effect of the electron-phonon
interaction on the spin texture of gapped Dirac helical fermions.
We will present for comparison, results for the familiar spectral
density ($A_{I}(k,\omega)$) associated with quasiparticle renormalization
which we compared with the much less familiar equivalent function
which enters gap renormalizations ($A_{z}(k,\omega)$). This is the
essential quantity for the calculation of the z-axis (perpendicular
to the surface states) properties. While only $A_{I}(k,\omega)$ is
needed in calculations of the density of states $N(\omega)$ both
$A_{z}$ and $A_{I}$ enter the dynamic conductivity $\sigma_{xx}(\omega)$.
In section II we present the necessary formalism for both the self
energy associated with quasiparticle renormalizations $\Sigma_{I}(\omega))$
and with the renormalization of the gap $\Sigma_{z}(\omega))$ which
determine the needed spectral functions $A_{I}(k,\omega)$ and $A_{z}(k,\omega)$.
Numerical results are presented in section III which deals with spectral
densities, density of states and optical conductivity. In section
IV we consider both in plane and out of plane spin texture. A summary
and conclusions are given in section V.

\section{Formalism}

\begin{figure}[tp]
\begin{centering}
\includegraphics[width=2.6in,height=3.2in,angle=-90]{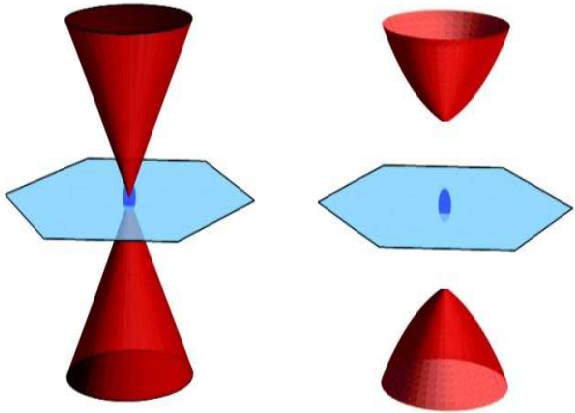} 
\par\end{centering}

\caption{(Color online) Illustrative representation of Dirac cones (left frame)
and including a gap $\Delta$ (right frame). }

\label{fig1} 
\end{figure}

\begin{figure}[tp]
\begin{centering}
\includegraphics[width=3.2in,height=3.2in]{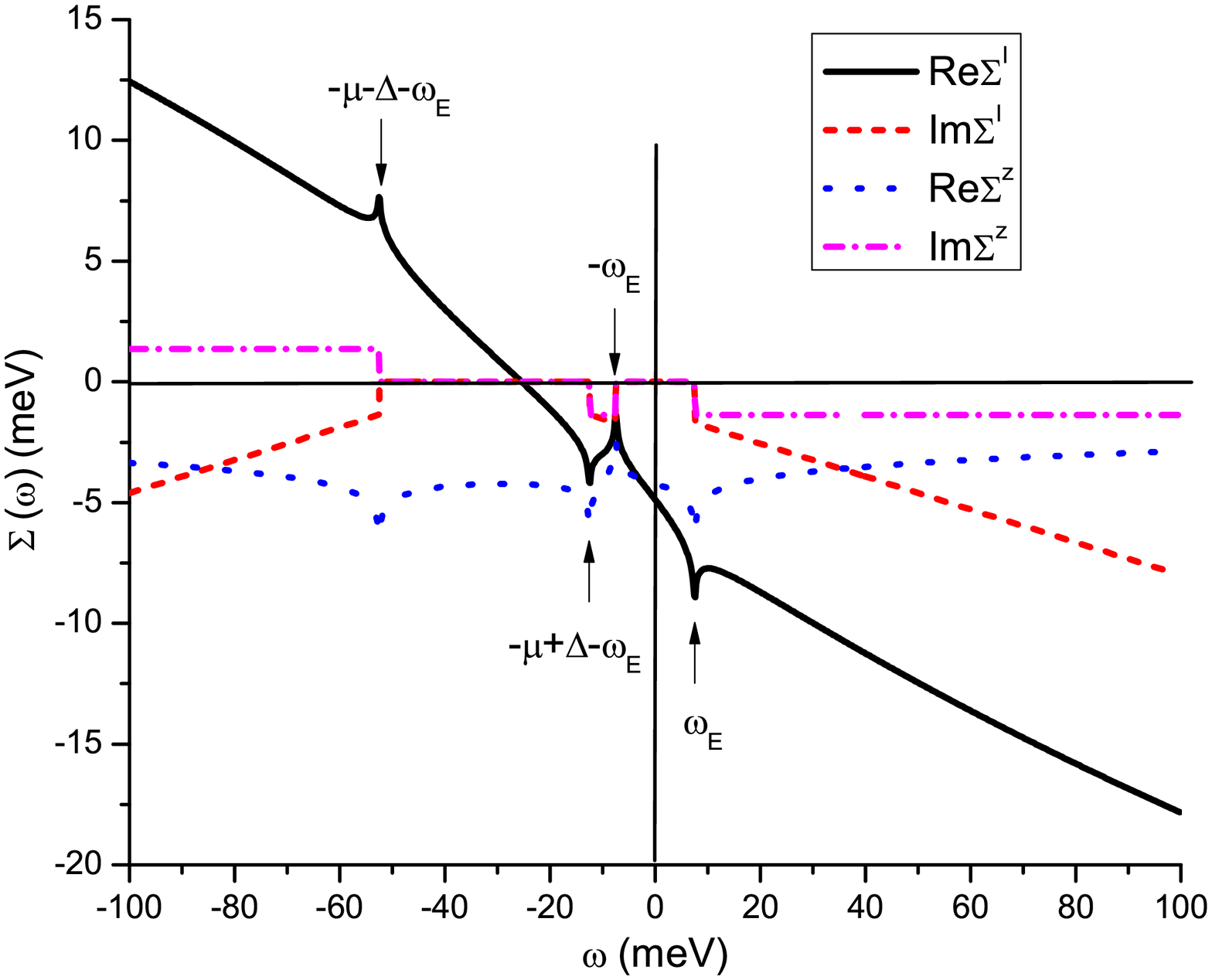} 
\par\end{centering}

\caption{(Color online) The real and imaginary part of the electron-phonon
self energies $\Sigma^{I}(\omega)$ and $\Sigma^{z}(\omega)$ in meV
as a function of $\omega$ in meV. $\Sigma^{I}(\omega)$ (real part,
solid black), and imaginary part (dashed red) gives the quasiparticle
renormalization and $\Sigma^{z}(\omega)$ (real part, dotted blue),
and imaginary part (dash-dotted purple) gives the gap renormalization.
Our choice of g=10 corresponds to a mass enhancement $\lambda\approxeq0.3$.}

\label{fig2} 
\end{figure}

We begin with a model hamiltonian for the electronic states at the
surface of a topological insulator which has the form

\begin{equation}
H_{0}=\hbar v_{F}[k_{x}\sigma_{y}-k_{y}\sigma_{x}]+\Delta\sigma_{z}+E_{0}(\mathbf{k})\label{Ham}
\end{equation}
where $\sigma$'s are Pauli spin matrices for real electron spin,
$v_{F}$ is the Fermi velocity of the Dirac electrons, $\mathbf{k}$
is momentum, $\Delta$ is a gap which can be introduced for example
by doping with magnetic impurities and $E_{0}(\mathbf{k})$ is a quadratic
term $E_{0}(\mathbf{k})=\frac{\hbar^{2}k^{2}}{2m}$ with $m$ the
electron mass. The introduction of $E_{0}$ gives particle hole asymmetry.
Here we will assume that $m$ is very large, as a first approximation,
and will ignore this term.

The non interacting Green's function takes on the form 
\begin{equation}
G_{0}(\mathbf{k},s,i\omega_{n})=\frac{1}{i\omega_{n}+\mu-s\sqrt{\hbar^{2}v_{F}^{2}k^{2}+\Delta^{2}}}
\end{equation}
with $i\omega_{n}$ the Matsubara frequencies and $\mu$ is the non
interacting chemical potential. The eigen energies are $\varepsilon_{k,s}=s\sqrt{\hbar^{2}v_{F}^{2}k^{2}+\Delta^{2}}$
with $s=\pm$ for conduction and valence band respectively. These
dispersion curves are shown schematically in Fig. 1 with (right frame)
and without (left frame) a gap.

The matrix ($2\times2$) Green's function $\hat{G}_{0}(\mathbf{k},i\omega_{n})$
can be written in terms of the scalar $G_{0}(\mathbf{k},s,i\omega_{n})$
as 
\begin{equation}
\hat{G}_{0}(\mathbf{k},i\omega_{n})=\frac{1}{2}\sum_{s=\pm}(1+s\mathbf{F}_\mathbf{k}\cdot\mathbf{\sigma})G_{0}(\mathbf{k},s,i\omega_{n})
\end{equation}
with the vector $\mathbf{F}_{k}$ defined as 
\begin{equation}
\mathbf{F}_\mathbf{k}=\frac{(-\hbar v_{F}k_{y},\hbar v_{F}k_{x},\Delta)}{\sqrt{\hbar^{2}v_{F}^{2}k^{2}+\Delta^{2}}}
\end{equation}
We want to include an electron-phonon interaction. The simplest is
the Holstein model for coupling to a phonon mode of energy $\omega_{E}$
with matrix element between electronic and phonon assumed to be a
constant ($g$). It is written as 
\begin{equation}
H_{e-ph}=-g\omega_{E}\sum_{\mathbf{k},\mathbf{k}^{\prime},s}c_{\mathbf{k},s}^{\dag}c_{\mathbf{k}^{\prime},s}(b_{\mathbf{k}^{\prime}-\mathbf{k}}^{\dag}+b_{\mathbf{k}-\mathbf{k}^{\prime}})\label{phonon}
\end{equation}
where $c_{\mathbf{k},s}^{\dag}$ creates an electron of momentum $\mathbf{k}$
and spin $s$ and $b_{\mathbf{k}^{\prime}-\mathbf{k}}^{\dag}$ creates
a phonon of momentum $\mathbf{k}^{\prime}-\mathbf{k}$ reduced to
the first surface state Brillouin zone and energy $\omega_{E}$ with
$g$ coupling constant. The perturbing Hamiltonian (5) provides a
self energy to electron motion which has two components one proportional
to the unit matrix $\widehat{I}$ and the other to $\widehat{\sigma}_{z}$.
Writing $\widehat{\Sigma}(i\omega_{n})=\Sigma^{I}(i\omega_{n})\widehat{I}+\Sigma^{Z}(i\omega_{n})\widehat{\sigma}_{z}$
we get in lowest order perturbation theory 
\begin{eqnarray}
 &  & \Sigma^{I}(i\omega_{n})=\frac{g^{2}\omega_{E}^{2}}{2}\sum_{\mathbf{q,}s}\notag\nonumber \\
 &  & \left[\frac{f_{F}(\varepsilon_{q,s})+f_{B}(\omega_{E})}{i\omega_{n}+\mu+\omega_{E}-\varepsilon_{q,s}}+\frac{f_{B}(\omega_{E})+1-f_{F}(\varepsilon_{q,s})}{i\omega_{n}+\mu-\omega_{E}-\varepsilon_{q,s}}\right]\label{sigmaI}
\end{eqnarray}
and 
\begin{eqnarray}
 &  & \Sigma^{Z}(i\omega_{n})=\frac{g^{2}\omega_{E}^{2}}{2}\sum_{\mathbf{q,}s}\frac{s\Delta}{\sqrt{\hbar^{2}v_{F}^{2}q^{2}+\Delta^{2}}}\times\notag\nonumber \\
 &  & \left[\frac{f_{F}(\varepsilon_{q,s})+f_{B}(\omega_{E})}{i\omega_{n}+\mu+\omega_{E}-\varepsilon_{q,s}}+\frac{f_{B}(\omega_{E})+1-f_{F}(\varepsilon_{q,s})}{i\omega_{n}+\mu-\omega_{E}-\varepsilon_{q,s}}\right]\label{sigmaZ}
\end{eqnarray}
where $f_{F}$ and $f_{B}$ are fermion and boson distribution functions
$1/[e^{(\omega-\mu)/k_{B}T}\pm1]$ respectively. Here $\mu$ is the
chemical potential which applies only to the electron distribution. Note that $\Sigma^{Z}(i\omega_{n})$ in Eq. (7)
is directly proportional to the gap $\Delta$ and will vanish for
$\Delta=0$. In terms of the self energies $\Sigma^{I}(i\omega_{n})$
and $\Sigma^{Z}(i\omega_{n})$ the interacting matrix Green's function
takes on the form 
\begin{equation}
\hat{G}(\mathbf{k},i\omega_{n})=\frac{1}{2}\sum_{s=\pm}(1+s\mathbf{H}_{\mathbf{k}}\cdot\mathbf{\sigma})G(k,s,i\omega_{n})\label{Greenf}
\end{equation}
with 
\begin{equation}
\mathbf{H}_{\mathbf{k}}=\frac{(-\hbar v_{F}k_{y},\hbar v_{F}k_{x},\Delta+\Sigma^{Z}(i\omega_{n}))}{\sqrt{\hbar^{2}v_{F}^{2}k^{2}+(\Delta+\Sigma^{Z}(i\omega_{n}))^{2}}}\label{Hk}
\end{equation}
and 
\begin{eqnarray}
&&G(k,s,i\omega_{n})=\nonumber\\
&&\frac{1}{i\omega_{n}+\mu-\Sigma^{I}(i\omega_{n})-s\sqrt{(\Delta+\Sigma^{Z}(i\omega_{n}))^{2}+\hbar^{2}v_{F}^{2}k^2}}\label{Green}
\end{eqnarray}
In Ref. (19) a factor $\Sigma^{Z\ast}(i\omega_{n})$ was mistakenly
introduced instead of $\Sigma^{Z}(i\omega_{n})$ in (9) and (10).
This leads to small numerical differences but has no qualitative significance.
The spectral function associated with $\widehat{I}$ and $\widehat{\sigma}_{z}$
matrix are 
\begin{eqnarray}
A_{I}(k,s,\omega) & = & -\frac{1}{\pi}ImG(k,s,i\omega_{n}\rightarrow\omega+i\delta)\nonumber \\
 & \equiv & A(k,s,\omega)
\end{eqnarray}
which is the quantity measured in angular resolved photo emission
spectroscopy ARPES.\cite{Zhou} Further 
\begin{eqnarray}
 &  & A_{z}(k,s,\omega)\notag\nonumber \\
 & = & -\frac{1}{\pi}ImTr[\sigma_{z}\hat{G}(k,s,i\omega_{n}\rightarrow\omega+i\delta)]\notag\nonumber \\
 & = & -\frac{1}{\pi}Im\{\frac{s[\Delta+\Sigma^{Z}(\omega+i\delta)]G(k,s,\omega+i\delta)}{\sqrt{\hbar^{2}v_{F}^{2}k^{2}+(\Delta+\Sigma^{Z}(\omega+i\delta))^{2}}}\}.
\end{eqnarray}
A considerable and instructive mathematical simplification of these
complicated expressions for (11) and (12) result when the imaginary
part of the z-component of the self energy $\Sigma^{Z}(i\omega_{n}\rightarrow\omega+i\delta)$
is ignored. We get 
\begin{equation}
A(k,s,\omega)=\frac{1}{\pi}\frac{Im\Sigma^{I}(\omega)}{[\widetilde{\omega}-s\sqrt{M}]^{2}+[Im\Sigma^{I}(\omega)]^{2}}
\end{equation}

\begin{eqnarray}
 &  & A_{z}(k,s,\omega)\notag\nonumber \\
 & = & \frac{s[\Delta+Re\Sigma^{Z}(\omega)]A(k,s,\omega)}{\sqrt{\hbar^{2}v_{F}^{2}k^{2}+(\Delta+Re\Sigma^{Z}(\omega))^{2}}}
\end{eqnarray}
with 
\begin{equation}
\widetilde{\omega}=\omega+\mu-Re\Sigma^{I}(\omega)
\end{equation}
and 
\begin{equation}
M=\hbar^{2}v_{F}^{2}k^{2}+(\Delta+Re\Sigma^{Z}(\omega))^{2}
\end{equation}
where $M$ is a function of $k$ and $\omega$. If we further take
the imaginary part of $\Sigma^{I}(\omega)$ to be zero (13) reduces
to a Dirac delta function but the real part of both $\Sigma^{I}$
and $\Sigma^{Z}$ remain which renormalize the single particle energies
and the gap. The probability of occupation of the state $\mathbf{k}$
denoted by $n(k)$ at temperature $T$ is given by\cite{Carbotte}
\begin{equation}
n(k)=\int_{-\infty}^{+\infty}d\omega A(k,s,\omega)f(\omega)
\end{equation}
and this is to be compared with the corresponding expression for the
z-component of spin $S_{z}(k)$ 
\begin{equation}
S_{z}(k)=\int_{-\infty}^{+\infty}d\omega A_{z}(k,s,\omega)f(\omega).
\end{equation}
We can also calculate the average value of the square root of the
sum of the squares of $x$ and $y$ component of spin which remains
locked perpendicular to momentum but its magnitude is changed by the
electron-phonon coupling 
\begin{equation}
\sqrt{S_{x}^{2}(k)+S_{y}^{2}(k)}=\int_{-\infty}^{+\infty}d\omega A_{x-y}(k,s,\omega)f(\omega)
\end{equation}
with 
\begin{equation}
A_{x-y}(k,s,\omega)=-\frac{1}{\pi}Im\{\frac{[\hbar v_{F}k]G(k,s,\omega)}{\sqrt{\hbar^{2}v_{F}^{2}k^{2}+(\Delta+\Sigma^{Z}(\omega))^{2}}}\}
\end{equation}
The density of electronic states $N(\omega)$ follows from $A(k,s,\omega)$
on integration over $k$ 
\begin{equation}
N(\omega)=\sum_{\mathbf{k,}s}A(k,s,\omega).\label{DOS}
\end{equation}
This quantity enters scanning tunneling microscopy experiments (STM).\cite{Miller,GLi,Nicol,Zhou,Pound}
The real part of the dynamic longitudinal conductivity which gives
the absorption spectrum for light follows as \cite{Li4} 
\begin{eqnarray}
 &  & Re\sigma_{xx}(\omega)=\frac{e^{2}v_{F}^{2}\pi^{2}}{\omega}\sum_{\mathbf{k}}\int_{-\infty}^{\infty}\frac{d\omega}{2\pi}[f(\omega)-f(\omega+\Omega)]\notag\nonumber \\
 &  & \times\lbrack A_{I}(\mathbf{k},\omega)A_{I}(\mathbf{k,}\omega+\Omega)-A_{z}(\mathbf{k},\omega)A_{z}(\mathbf{k,}\omega+\Omega)]
\end{eqnarray}
with $A_{I}(\mathbf{k},\omega)=\sum_{s}A_{I}(k,s,\omega)$ and $A_{z}(\mathbf{k},\omega)=\sum_{s}A_{z}(k,s,\omega)$.
Note that both $A_{z}$ and $A_{I}$ enter this quantity even though
it is an in-plane property.

\section{Numerical results for self energy and derived quantities}

\begin{figure}[tp]
\begin{centering}
\includegraphics[width=3.2in,height=3.2in]{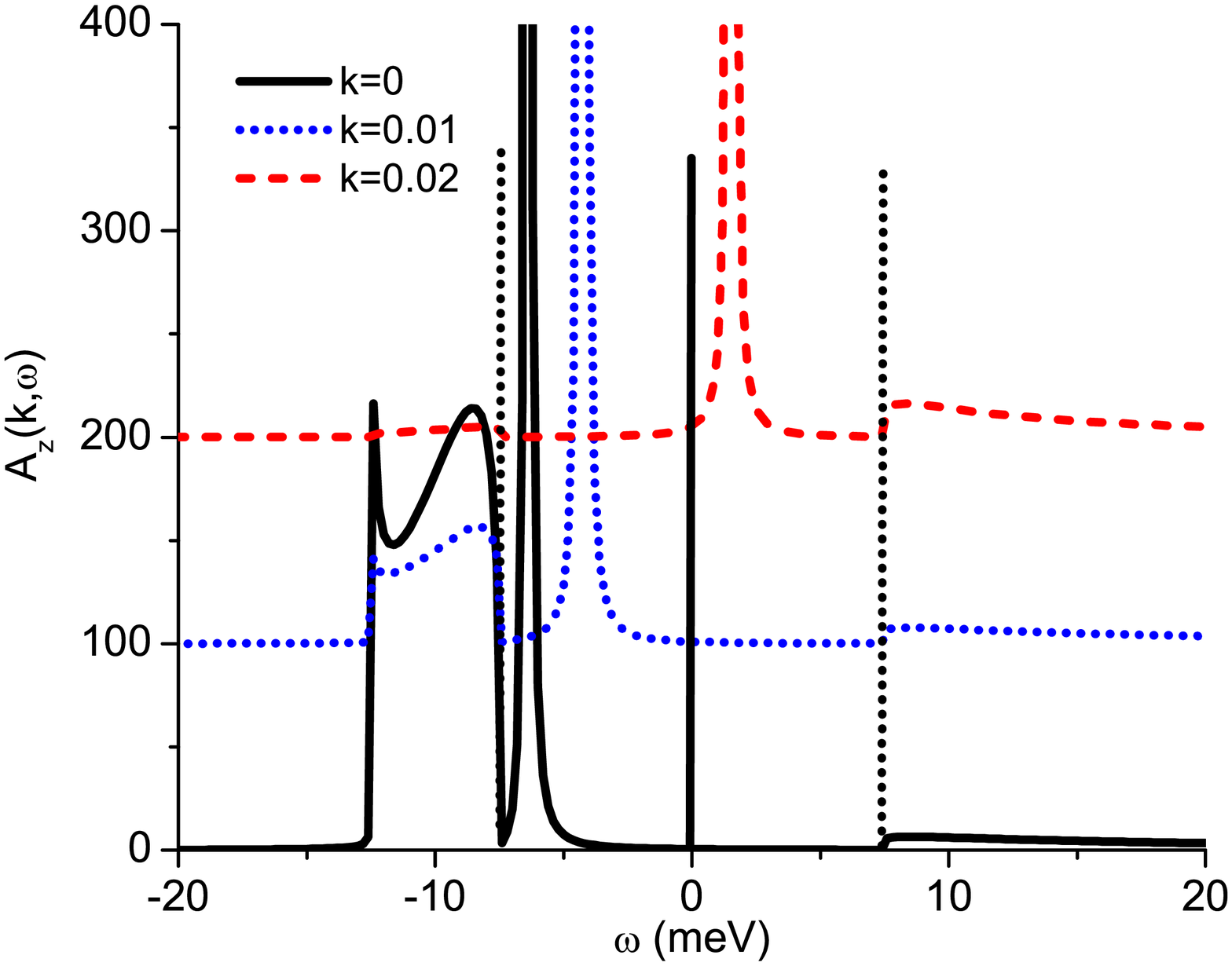} \includegraphics[width=3.2in,height=3.2in]{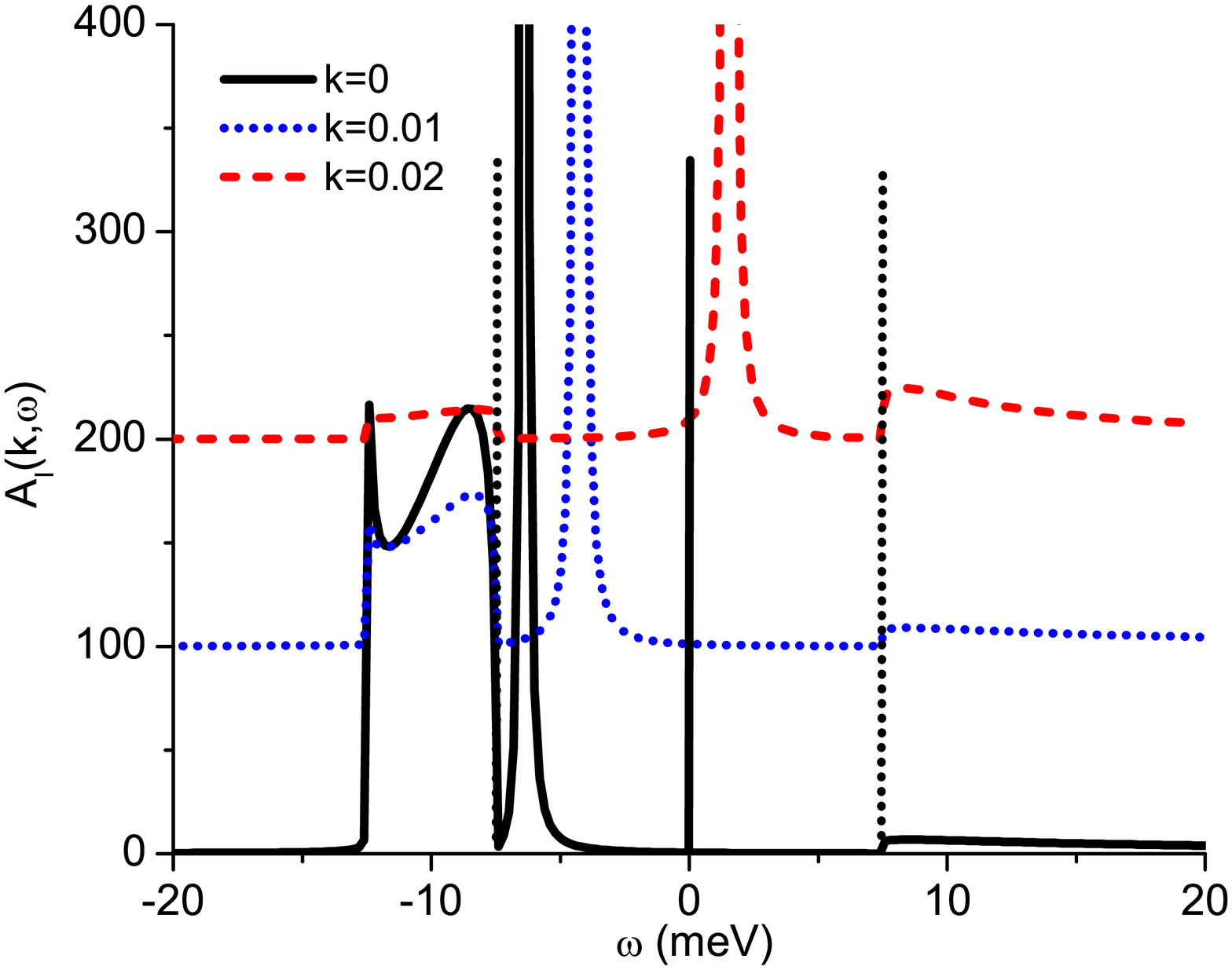} 
\par\end{centering}

\caption{(Color online) The spectral density $A(k,\omega)$ as a function of
energy $\omega$ in meV for three values of momentum $k$ namely $k=0$
(solid black), $k=0.01$ (dotted blue) and $k=0.02$ (dashed red)
in units of the inverse lattice parameter $a$. The top frame is for
$A_{z}(k,\omega)$ (associated with the gap renormalization) and the
bottom frame is for $A_{I}(k,\omega)$ associated with quasiparticle
renormalization. Vertical dotted black lines are at $\omega=\pm\omega_{E}$
with phonon Einstein energy set at 7.5 meV. The gap $\Delta=20$ meV and the bare chemical potential
is $\mu=25$ meV.}

\label{fig3} 
\end{figure}

\begin{figure}[tp]
\begin{centering}
\includegraphics[width=3.3in,height=3.3in,angle=-90]{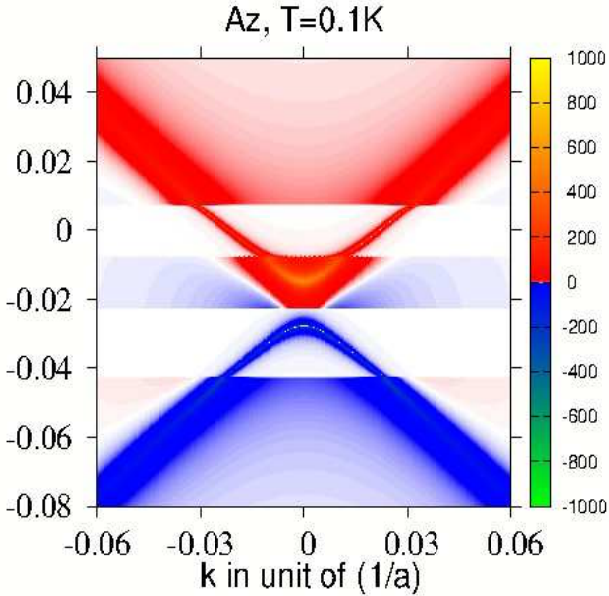} 
\par\end{centering}

\caption{(Color online) A color plot of the gap spectral density $A_{z}(k,\omega)$
as a function of energy $\omega$ in (eV) vertical axis and momentum
$k$ in ($1/a$) horizontal axis, with $a$ the lattice parameter.
The gap $\Delta=10$ meV and the Einstein phonon energy $\omega_{E}=7.5$
meV. }
\label{fig4} 
\end{figure}

\begin{figure}[tp]
\begin{centering}
\includegraphics[width=3in,height=3.2in]{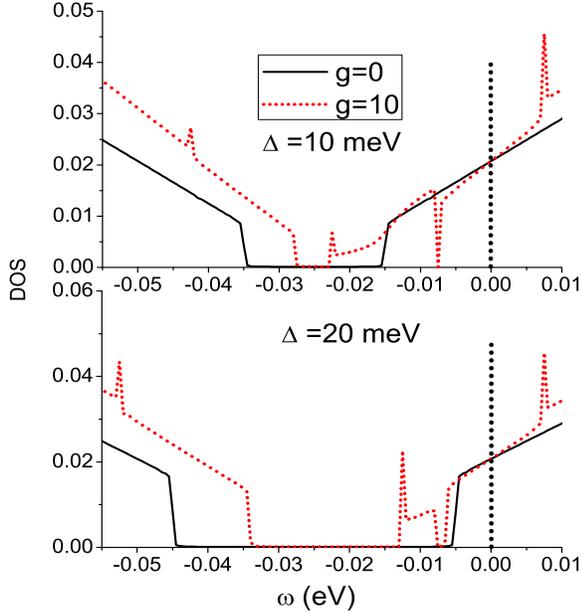} 
\par\end{centering}
\caption{(Color online) The density of states $N(\omega)$ in inverse eV versus
energy $\omega$ in eV. The dotted red line includes the electron-phonon
interaction with $g=10$ in Eq. (6)-(7). The solid black line which
is for comparison is for bare bands ($g=0$). The top frame has a
gap $\Delta=10$ meV while the bottom frame has a gap of 20 meV. }
\label{fig5} 
\end{figure}

\begin{figure}[tp]
\begin{centering}
\includegraphics[width=3.2in,height=3.2in]{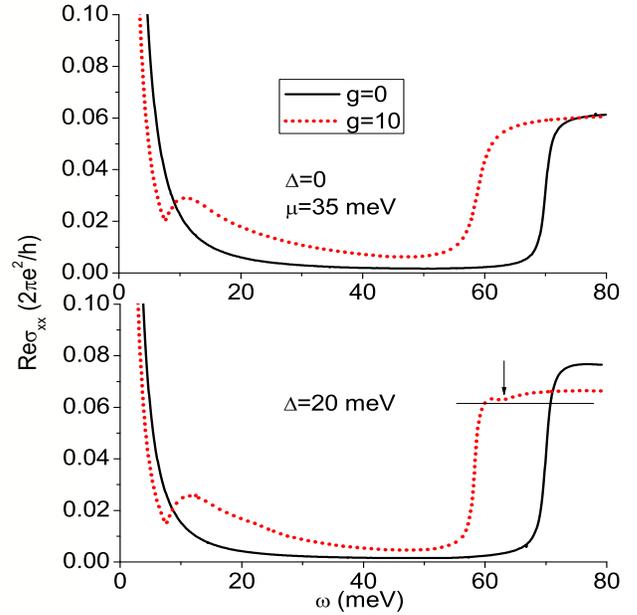} 
\par\end{centering}

\caption{(Color online) The real part of the longitudinal dynamic optical conductivity
$Re\sigma_{xx}(\omega)$ in units of $e^{2}/\hbar$ as a function
of photon energy $\omega$ in meV. The dotted red line includes the
electron-phonon interaction with $g=10$ in Eq. (6)-(7). The solid
black line, which is for comparison, is for bare bands ($g=0$). The
top frame has no gap $\Delta=0$ and the bottom frame includes a gap
$\Delta=20$ meV. In both cases the bare chemical potential $\mu=35$
meV. The arrow identifies the phonon structure.}

\label{fig6} 
\end{figure}

In lowest order perturbation theory the chemical potential $\mu$
in Eq. (6) and (7) is to be interpreted as having its non interacting,
independent particle value and will be quoted in all the results to
be presented in this paper. With electron-phonon coupling the chemical
potential will in general be shifted from its free band value to $\mu_{interacting}=\mu+Re\Sigma^{I}(i\omega_{n}\rightarrow\omega+i\delta)$.
In Fig. 2 we show our results for $\Sigma(i\omega_{n}\rightarrow\omega+i\delta)$
in meV as a function of energy $\omega$ in meV where we have analytically
continued from Matsubara to real frequencies. The solid black curve
is $Re\Sigma^{I}$ with $Im\Sigma^{I}$ given by the dashed
red curve while the dotted blue is for $Re\Sigma^{Z}(\omega)$
with its imaginary part $Im\Sigma^{Z}(\omega)$ given by the purple
dash-dotted curve. First the real part of
the quasiparticle self energy is not zero at $\omega\simeq0$ but
rather is of the order of $\simeq-5$meV which reduces the chemical
potential of the interacting picture by about 1/4 of its bare band
value here taken to be 25 meV with the gap $\Delta$ equal to 20 meV.
With $g$ set to a value of 10 in Eq. (6) and (7), the mass enhancement parameter $\lambda\sim$ 0.3 as can be seen 
in Fig. 2. The parameter $\lambda$ is the slope of the real part of $\Sigma^I(\omega)$ at $\omega=0$, i.e. $\lambda=(d/d\omega)Re\Sigma^I(\omega)$
at $\omega=0$. We have also taken the Einstein phonon energy to be 7.5 meV which is representative of what is seen in many
but not all experiments in $Bi_{2}Se_3$ and has been identified in the work of X. Zhu et al. \cite{Zhu1} as a surface phonon.
As discussed in the introduction coupling to specific optical phonons have been found in other experiments \cite{Zhu2,Pan,Hatch,Chen,LaForge}.
Of course there can also be coupling to acoustic phonons as studied by Giraud et al..\cite{Giraud1,Giraud2} 
These authors use a continuum model for the phonons and a deformation potential model to describe their coupling 
to the helical surface Dirac fermions with due attention to the modifications brought about by helicity of the charge carriers. 
In our simplified model these complications enter only in determining the size of $g$ in Eq. (6) and (7) which we set 
through consideration of experiment (i.e. $\lambda\sim$ 0.3). In principle these could also be coupling to electronic modes 
but this is not considered in this work which deals with electron-phonon coupling alone. Our choice of gap $\Delta=10$ mev is representative
of the work of Chen et al. \cite{Chen2} who found in ARPES
experiments a gap of $\simeq-7$ mev in $Bi_{2}Se_3$ magnetically dopped with
$Mn$. We note prominent phonon structure at $\omega=\pm\omega_{E}=\pm7.5$
meV by choice, with additional structures at $-\mu+\Delta-\omega_{E}$
and at $-\mu-\Delta-\omega_{E}$. For $\omega>0$ the imaginary
part of the quasiparticle self energy is zero in the interval ($0,\omega_{E}$)
after which its absolute magnitude is finite and increases linearly
with increasing value of $\omega$. This dependence reflects the linear
dependence on energy of the underlying bare density of states. Also
it needs to be negative because $-Im\Sigma^{I}(i\omega_{n}\rightarrow\omega+i\delta)$
is a scattering rate, and hence positive. By contrast the imaginary part of the gap self energy is flat because the extra factor of 
$\Delta/\sqrt{\hbar^2v_F^2q^2+\Delta^2}$ appearing in (7) but not in (6) compensates for the density of states variarion which comes from the sum 
over $q$. Note also that both self energies have a finite imaginary part in the interval $\omega\in(0,-\omega_E)$ and $\omega<-\mu-\Delta-\omega_E$. 
While the real part of the quasiparticle self energy can be both positive or negative, the real
part of the gap self energy is everywhere negative and so decreases
the bare gap $\Delta$ at all $\omega$ considered here. Its magnitude
is everywhere of order 1/4 of the input value of $\Delta$ with some
$\omega$ dependence and is encoded with boson structure. These self
energies have a profound effect on the corresponding spectral densities
of Eq. (11) and (12) as we see in Fig. 3 where the top frame applies
to $A_{z}(k,\omega)=A_{z}(k,s>0,\omega)$ (conduction band) and the
bottom frame is for $A_{I}(k,\omega)=A_{I}(k,s>0,\omega)$. For the
range of values of $k$ shown namely, $k=0.0$, $0.01$ and $0.02$
in units of the inverse of the lattice parameter ($a$), these two
functions do not differ much from each other. This is traced to the
fact that $A_{z}(k,\omega)$ has the extra factor $\frac{\Delta}{\sqrt{\hbar^{2}v_{F}^{2}k^{2}+\Delta^{2}}}$
as compared with $A_{I}(k,\omega)$ where we have, for simplicity, neglected renormalizations as these do not change the argument in an important
way. As long as $\hbar v_{F}k$ is small compared with $\Delta$ the
relevant additional factor is near one. For large $k$ however, larger than
those shown in Fig. 3, $A_{z}(k,\omega)$ will become much smaller
in magnitude than $A_{I}(k,\omega)$ by a factor of $\Delta/(\hbar v_{F}k)$.
Returning to the top frame of Fig. 3 we first note in the solid black
curve for $k=0$ (which is the momentum of the bottom of the bare
conduction band) a large, only slightly broaden and shifted in energy,
quasiparticle peak centered approximately at $\simeq6$ meV below
the Fermi surface at $\omega\simeq0$. For the bare bands it would
be at $5$ meV instead. But with correlations there are also additional
features. There is a large increase in $A_{z}(k,\omega)$ starting
at $\omega=-\omega_{E}=-7.5$meV followed by a large boson structure
extending down to $\simeq-12.5$meV where it drops to zero. Such a
feature is also seen in the dashed red curve but it is much smaller.
The lower cut off is identified to correspond to the energy $-\mu-\omega_{E}+\Delta$.
If we had shown the valence band contribution to the spectral density
it would have a further boson structure at $-\mu-\omega_{E}-\Delta$
as in the self energy of Fig. 2. It is clear from this description
that the effect of the electron-phonon interaction on the bare band
is much more complicated than a simple constant shift in gap value
and a slight shift in quasiparticle energies with small broadening.
Note also the clear phonon sidebands at energies above $\omega=\omega_{E}$
in all three curves. The black vertical dotted lines identify energies
$\omega=\pm\omega_{E}$ for easy reference. A different representation
of these changes is given in Fig. 4 where we show a false color plot
of $A_{z}(k,\omega)$ as a function for energy $\omega$ along the
vertical axis and momentum $k$ along the horizontal. As seen in Eq. (12) and (14), $A_{z}$ carries the sign of $s$ and is 
positive (red) in the renormalized conduction band and negative (blue) in the renormalized valence band. The phonon at
$\omega=\pm\omega_{E}$ is clearly identified as is the bottom of
the renormalized conduction band at $-\mu-\omega_{E}+\Delta$ and the valence band
phonon structure at $-\mu-\omega_{E}-\Delta$. Particularly striking
is the large modification of the bare band dispersion curves in the
region just above the renormalized conduction band minimum.

The large renormalization effects seen in Fig. 3 and Fig. 4 can have
a profound effect on certain quantities while at the same time have
much more modest manifestations in others as we will now describe.
In Fig. 5 we show results for the density of states (DOS) $N(\omega)$
given in Eq. (21). The top frame is for a gap $\Delta=10$ meV and
the bottom for $\Delta=20$ meV. The red dotted curve is the renormalized
quasiparticle DOS while the black solid curve is the bare band case
shown for comparison. In both frames the bare chemical potential is
set at $25$ meV, which is $5$ meV above the gap in the conduction
band (for the case $\Delta=20$ meV). The energy variable $\omega$ is set so that
$\omega=0$ corresponds to the Fermi surface so that the bottom of
the conduction band is at $-5$ meV in bottom frame and $-15$ meV
in top frame with the top of the valence band at $-45$ meV and $-35$
meV respectively. In both cases the remaining gap between valence
and conduction band has been very much reduced over its bare band
value. The four phonon structures at $\omega=\pm\omega_{E},-\mu-\omega_{E}-\Delta$
and $-\mu-\omega_{E}+\Delta$ are clearly seen with the bottom of
the conduction band given by the energy $-\mu-\omega_{E}+\Delta$.
By comparison, the corresponding boson structures in the real part
of the dynamic longitudinal optical conductivity $Re\sigma_{xx}(\omega)$
which is more closely related to a convolution of two DOS factors and
is given in Eq. (22), are much more modest as seen in Fig. 6. The top
frame is for gapless Dirac fermions ($\Delta=0$) and is included
for comparison while the bottom frame is for the gapped case with
$\Delta=20$ meV. The solid black curves are for the bare band case
with a small residual scattering rate included to broaden out the
Drude peak due to intraband transitions. This peak is large only at small
$\omega$ and is centered at $\omega=0$. The bare chemical potential
is $35$ meV and we see the onset of a second absorption
band coming from the interband transitions which start at $\omega=2\mu$.
These transitions continue up to large energies and provide the so
called universal background. In our units for $Re\sigma_{xx}(\omega)$
which is $e^{2}/\hbar$, this background\cite{Gusynin1,Gusynin2}
has a height of $1/16$. As is clear in the top frame for $\Delta=0$
this height is almost unaffected by the electron-phonon interaction
(red dotted curve). Staying with the top frame we see however the
appearance of the Holstein processes above $\omega=\omega_{E}$ which
provides significant phonon assisted absorption in the photon region
above $\omega_{E}$ and below the main interband absorption edge at
$2\mu$. The other feature to be noted is that, for the correlated
case, the onset of the interband transitions has moved to lower energies
and is now at twice the value of the interacting chemical potential.
The lower frame which is for the gap fermion case has another important
element. As is well known,\cite{Stille,Gusynin1,Gusynin2} when $\Delta\neq0$
there is a peak in the interband transitions just above the threshold
energy which persists up to a few $\Delta$ above the threshold before
the value of the background is reestablished at its universal value.
In the clean limit we have the analytic result 
\begin{eqnarray}
Re\sigma_{xx}(\omega) & = & \frac{e^{2}}{4\hbar}[\delta(\omega)\frac{\mu^{2}-\Delta^{2}}{|\mu|}\theta(\mu^{2}-\Delta^{2})+\notag\nonumber \\
 &  & \frac{\omega^{2}+4\Delta^{2}}{4\omega^{2}}\theta(\omega-2\max(|\mu|,\Delta))]
\end{eqnarray}
where $\theta$ is a heaviside function. We see that, just above the
interband onset the conductivity is larger than its universal value
of $\frac{e^{2}}{16\hbar}$ as is seen most clearly in the solid black curve of
the lower frame of Fig. 6 which is the bare band result. The horizontal
straight line segment indicate $1/16$. We note that, with electron-phonon
(dotted red curve), the magnitude of the absorption in the region of
the edge still remains above the universal background value but now
there is also a small phonon structure highlighted by the vertical
arrow. No such structure is seen in the top frame for $\Delta=0$.
It is the variation with energy of the background (in the presence
of a finite gap) which allows for the phonon structure to be revealed.

\section{Spin texture}

\begin{figure}[tp]
\begin{centering}
\includegraphics[width=3.2in,height=3.2in]{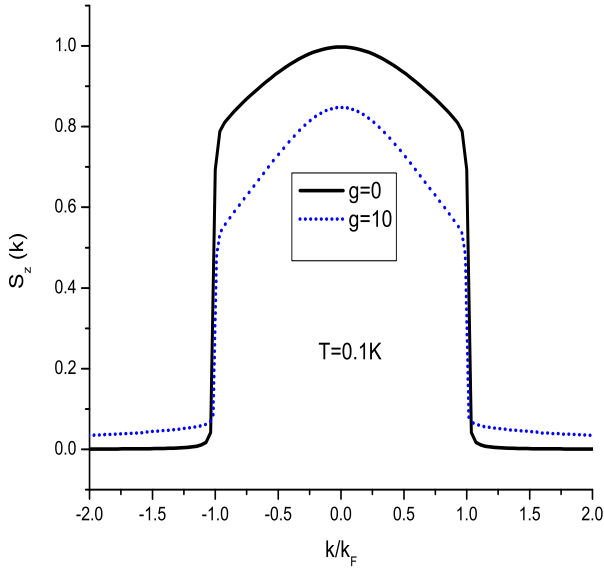} 
\par\end{centering}

\caption{(Color online)The z-component of spin $S_{z}(k)$ as a function of
momentum $k$ normalized to $k_{F}$. The solid black line is for
$g=0$ (no electron-phonon interaction) and is for comparison with
the dotted blue curve which includes the self energy with coupling
$g=10$ (Eq. (6)-(7)). }

\label{fig7} 
\end{figure}

\begin{figure}[tp]
\begin{centering}
\includegraphics[width=3.2in,height=3.2in]{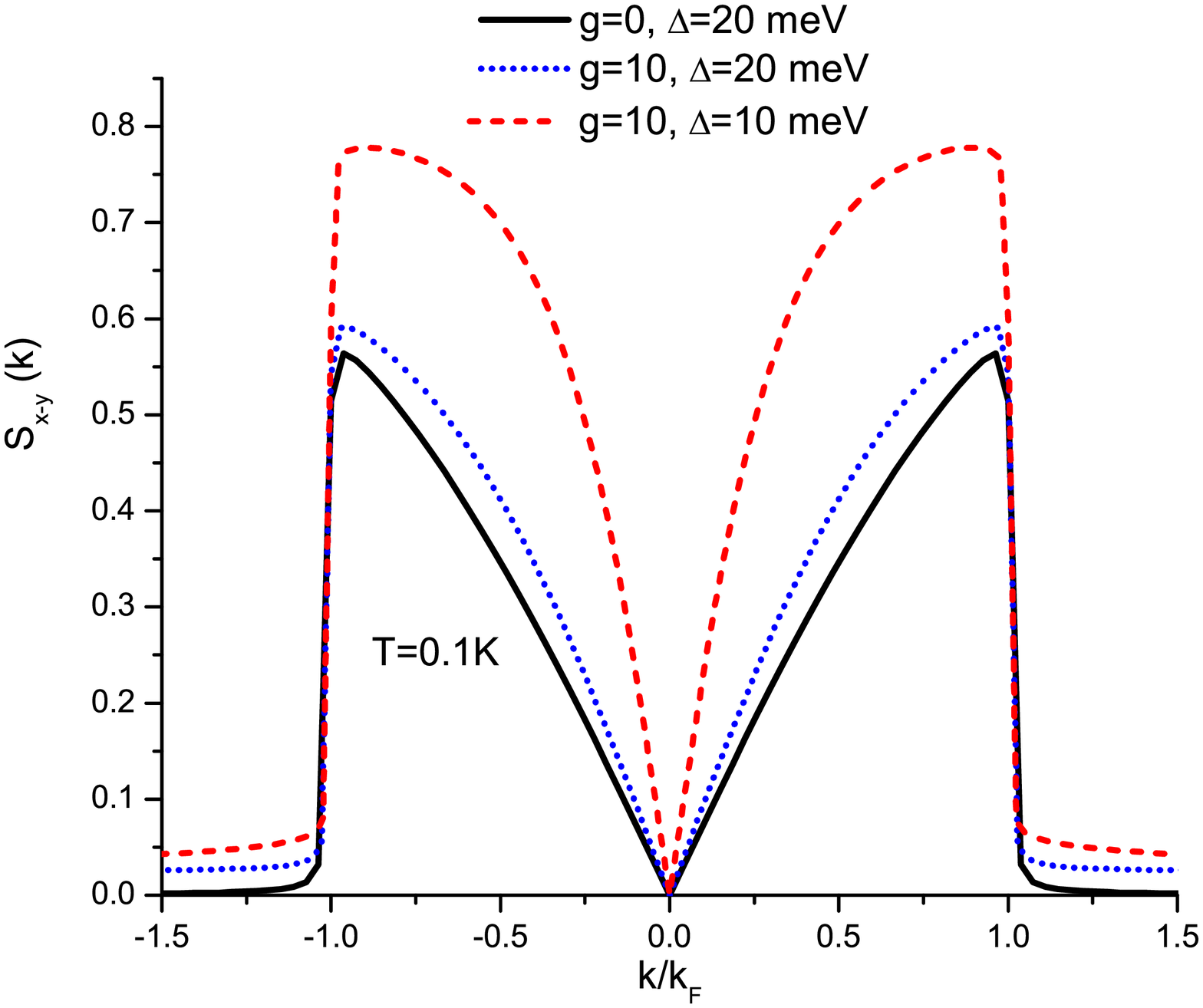} 
\par\end{centering}

\caption{(Color online) The magnitude of the in plane spin $S_{x-y}(k)$ as
a function of momentum $k$ normalized to its value at the Fermi surface
$k_{F}$. The solid black curve is for comparison and involves the
bare bands ($g=0$) with gap $\Delta=20$ meV. The dotted blue has
the same gap but includes the electron-phonon renormalizations with
coupling $g=10$ in Eq. (6)-(7). The dashed red is for a smaller value
of gap $\Delta=10$ meV. }

\label{fig8} 
\end{figure}

\begin{figure}[tp]
\begin{centering}
\includegraphics[width=3.2in,height=3.2in]{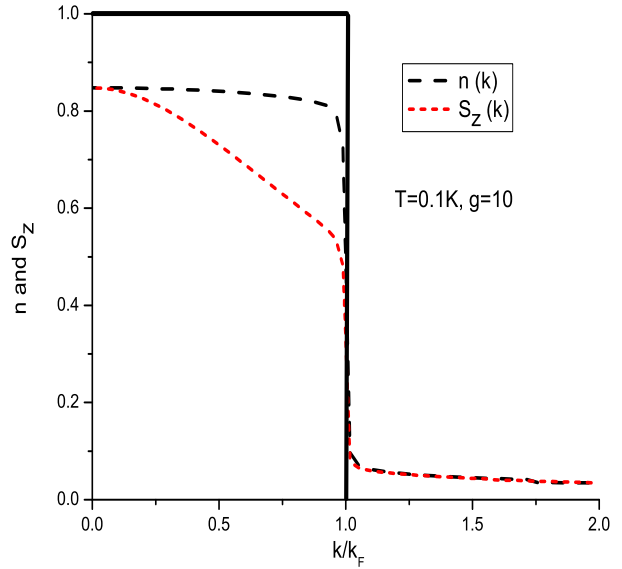} 
\par\end{centering}

\caption{(Color online) Comparison of the z-component of spin $S_{z}(k)$ of
Eq. (18) as a function of momentum $k$ normalized to its value at
the Fermi surface $k_{F}$ (short dashed red curve) with the probability
of occupation of state $n(k)$ of Eq. (17) (long dashed black curve).
The heavy solid black line gives the case of bare bands for comparison. }

\label{fig9} 
\end{figure}

The z-axis spin texture is determined from the spectral density
$A_{z}(k,s,\omega)$ of Eq. (12). Results are presented in Fig. 7
for $S_{z}(k)$ as a function of momentum $k$ normalized to its value
at the Fermi surface ($k_{F}$). $S_{z}(k)$ involves an integral
over frequency $\omega$ of the overlap of $A_{z}(k,s,\omega)$ and
the Fermi function $f(\omega)$ as given in Eq. (18). We only show
results for the conduction band $s=+$. For the valence band $s=-1$
and $S_{z}(k)$ will change sign. The gap $\Delta=20$ meV and the
chemical potential $\mu=25$ meV as before in section III. The solid
black curve is the bare band result for temperature $T=0.1K$. We
see an abrupt drop to zero at $k/k_{F}=1$ because of the thermal
factor which here is basically a Heaviside function with cut off at
the Fermi surface. By contrast the dotted blue curve includes the
electron-phonon renormalization discussed at length in the previous
section. Now the magnitude of $S_{z}(k)$ at $k=0$ is reduced below
its bare band value. More importantly the jump at $k/k_{F}$ is smaller than it is in the solid curve and further, there are finite
tails beyond this momentum which are entirely due to many body correlation
effects that go beyond a bare band description. They can be taken
as representative of other correlation effect such as those due to
electron-electron interactions rather than electron-phonon. Compared
to the modifications found in the spectral densities, the density
of states and the conductivity, these effects are certainly not as
spectacular. Mathematically this results from the fact that $S_{z}(k)$
involves an integration over energy $\omega$ while it is this energy
dependence which reflects most directly the details of the correlation
effects.

The in-plane spin texture is also changed, although the perpendicular
spin momentum locking remains. In Fig. 8 we show results for the momentum
dependence (in units of $k/k_{F}$) of $S_{x-y}(k)$ defined in Eq.
(19). It gives the magnitude of the in-plane spin at momentum $k$.
Without interactions it would be equal to $\frac{\hbar v_{F}k}{\sqrt{\hbar^{2}v_{F}^{2}k^{2}+\Delta^{2}}}\theta(k_{F}-k)$
and is shown as the solid black curve for a gap of $20$ meV and temperature
$T=0.1$K. It starts at zero for $k=0$ and rises monotonically with
a sharp cut off at $k/k_{F}=1$ at zero temperature. When the electron-phonon
interaction is included through Eq. (20) we get the dotted blue
curve which follows closely the solid black curve with the important
difference that it has tails beyond $k=k_{F}$ characteristic of correlation
effects. The dashed red curve is for the same value of $g$
but now the gap has been reduced to $10$ meV. In this case the curve
rises more sharply out of $k=0$, flattens before showing a sharp
drop at $k=k_{F}$ and the correlation tails beyond this drop off
are now considerably larger. The correlation tails seen in $S_{z}(k)$
as well as in $S_{x-y}(k)$ are very similar to the much more familiar
case\cite{Carbotte} of the momentum distribution in a fermion system
with electron-phonon interaction. The probability of occupation of
the state $k$ at zero temperature denoted by $n(k)$ is given in
Eq. (17) and in the bare band picture is a step function $\theta(k_{F}-k)$.
As shown in Fig. 9 long dashed black curve for $T=0.1$K, $g=10$
and a gap $\Delta=20$ meV, $n(k)$ is considerably reduced from value
one throughout the occupied states. It still has a finite discontinuous
jump at $k=k_{F}$ with correlation tail for $k>k_{F}$. The short
red dashed curve is $S_{z}(k)$ previously calculated and presented
here for comparison with $n(k)$. This shows the qualitatively similar
effect of the electron-phonon interaction on these two quantities.

\section{Conclusions}

We calculated the effect of electron-phonon coupling on a system of
helical gapped Dirac fermions. A simple Holstein model with coupling
to a single Einstein phonon of energy $\omega_{E}$ was used. For
massive Dirac fermions two self energies need to be introduced. There
is the familiar quasiparticle renormalizarion which changes the bare
band energies to dressed dispersion curves and provides damping. In
addition there is a second self energy directly associated with modifications
of the gap. It is energy ($\omega$) dependent and complex with the
real part directly modulating the magnitude of the bare band gap.
Both these self energies enter the quasipaticle spectral density $A(k,s,\omega)$
with $k$ momentum $\omega$ energy and $s=\pm$ giving conduction
and valence band respectively. In direct analogy, a second spectral
density $A_{z}(k,s,\omega)$ associated with the gap channel is also
introduced. The first function $A(k,s,\omega)$ can be measured directly
in angular resolved photo-emission spectroscopy (ARPES). Its average
over momentum $k$ determines the electronic density of states $N(\omega)$
measured in scanning tunneling spectroscopy (STM). The second $A_{z}(k,s,\omega)$
enters along with $A(k,s,\omega)$ the expressions for the dynamic
longitudinal optical conductivity. It also determines the out of plane
spin texture. Both $A$ and $A_{z}$ are functions of $\omega$ and
are encoded with sharp phonon structures at $\omega=\pm\omega_{E},-\mu-\omega_{E}-\Delta$
and $-\mu-\omega_{E}+\Delta$. These phonon structures manifest directly
in the density of states and in a somewhat different way in the optical
conductivity.

The out of plane and in plane spin textures are modified by the electron-phonon
interaction. But these quantities involve an integration over energy
of an overlap of spectral density and thermal factor. Consequently
the manifestation of phonon-electron coupling in these quantities
is more subtle and not as direct. Nevertheless important corrections
to a bare band picture arise. Just as the probability of occupation
of a state of momentum $k$ is reduced from one for $k$ less than
$k_{F}$, and extended (in momentum) tails appear for $k>k_{F}$,
the magnitude of $S_{z}(k)$ follows the same trend. A similar picture
applies for the magnitude of the in plane spin component. While the
spin remains locked in the direction perpendicular to momentum, correlation
tails appear beyond $k=k_{F}$. While the calculations are for the explicit
case of the electron-phonon interaction, they serve to illustrate
how many body interactions in general modify the bare band picture.
\begin{acknowledgments} This work was supported by the Natural Sciences
and Engineering Research Council of Canada (NSERC) and the Canadian
Institute for Advanced Research (CIFAR). This material is also based upon 
work supported by the National Science Foundation under the NSF EPSCoR Cooperative 
Agreement No. EPS-1003897 with additional support from the Louisiana Board of Regents.
JPC initiated the work and wrote much of the text, ZL did
all the calculations.
\end{acknowledgments}

\end{document}